\documentclass[longauth,breaklinks=true]{aa}

\usepackage{silence}
\usepackage{graphicx}
\usepackage{lscape}
\usepackage{txfonts}
\usepackage{stackengine}
\usepackage{subcaption}
\usepackage{geometry}
\usepackage{float}
\usepackage{tabularx}
\usepackage{longtable}
\usepackage{booktabs}
\usepackage{gensymb}
\usepackage{textgreek}
\usepackage{textcomp}
\usepackage{multirow}
\usepackage{orcidlink}

\usepackage{natbib,twoopt}

\bibpunct{(}{)}{;}{a}{}{,}
\hypersetup{
    colorlinks=true,
    linkcolor=red,
    citecolor=blue,
    urlcolor=blue
}

\begin{document}

   \title{Unveiling blazar synchrotron emission: a multiwavelength polarimetric study of HSP and LSP populations}

   \titlerunning{Probing blazar emission through polarimetric measurements}
    \authorrunning{S. Capecchiacci et al.}

\author{
  Sara Capecchiacci \inst{\ref*{forth}, \ref*{physics_crete}}\orcidlink{0009-0007-1918-577X} \and
  Ioannis Liodakis \inst{\ref*{forth}, \ref*{nasa_marshall}}\orcidlink{0000-0001-9200-4006} \and
  Riccardo Middei\inst{\ref*{inaf_obs_rome}, \ref*{asi_datacenter}, \ref*{harv_smithsonian}}\orcidlink{0000-0001-9815-9092} \and
  Dawoon E. Kim\inst{\ref*{inaf_rome}} \orcidlink{0000-0001-5717-3736} \and
  Laura Di Gesu\inst{\ref*{asi}}\orcidlink{0000-0002-5614-5028} \and
  Iv\'{a}n Agudo\inst{\ref*{granada}}\orcidlink{0000-0002-3777-6182} \and
  Beatriz Ag\'{i}s-Gonz\'{a}lez \inst{\ref*{forth}}\orcidlink{0000-0001-7702-8931} \and
  Axel Arbet-Engels\inst{\ref*{planck_garching}}\orcidlink{0000-0001-9076-9582} \and
  Dmitry Blinov \inst{\ref*{forth}, \ref*{physics_crete}} \and
  Chien-Ting Chen\inst{\ref*{huntsville}}\orcidlink{0000-0002-4945-5079} \and
  Steven R. Ehlert\inst{\ref*{nasa_marshall}}\orcidlink{0000-0003-4420-2838} \and
  Ephraim Gau\inst{\ref*{physics_washington}}\orcidlink{0000-0002-5250-2710} \and
  Lea Heckmann\inst{\ref*{planck_garching}, \ref*{paris}}\orcidlink{0000-0002-6653-8407} \and
  Kun Hu\inst{\ref*{physics_washington}} \and
  Svetlana G. Jorstad\inst{\ref*{boston},\ref*{st.petersb}}\orcidlink{0000-0001-6158-1708} \and
  Philip Kaaret\inst{\ref*{nasa_marshall}}\orcidlink{0000-0002-3638-0637} \and
  Pouya M. Kouch\inst{\ref*{physics_finland},\ref*{finnish_eso}}\orcidlink{0000-0002-9328-2750} \and
  Henric Krawczynski\inst{\ref*{physics_washington}}\orcidlink{0000-0002-1084-6507} \and
  Elina Lindfors\inst{\ref*{physics_finland}}\orcidlink{0000-0002-9155-6199} \and
  Fr\'{e}d\'{e}ric Marin\inst{\ref*{strasbourg}}\orcidlink{0000-0003-4952-0835} \and
  Alan P. Marscher\inst{\ref*{boston}}\orcidlink{0000-0001-7396-3332} \and
  Ioannis Myserlis\inst{\ref*{mm_granada}}\orcidlink{0000-0003-3025-9497} \and
  Stephen L. O'Dell\inst{\ref*{nasa_marshall}}\orcidlink{0000-0002-1868-8056} \and
  Luigi Pacciani\inst{\ref*{inaf_rome}}\orcidlink{0000-0001-6897-5996} \and
  David Paneque\inst{\ref*{planck_garching}} \and
  Matteo Perri\inst{\ref*{asi_datacenter},\ref*{inaf_obs_rome}}\orcidlink{0000-0003-3613-4409} \and
  Simonetta Puccetti\inst{\ref*{asi_datacenter}}\orcidlink{0000-0002-2734-7835} \and
  M. Lynne Saade\inst{\ref*{huntsville},\ref*{nasa_marshall}}\orcidlink{0000-0001-7163-7015} \and
  Fabrizio Tavecchio\inst{\ref*{inaf_merate}}\orcidlink{0000-0003-0256-0995} \and
  Allyn F. Tennant\inst{\ref*{nasa_marshall}}\orcidlink{0000-0002-9443-6774} \and
  Efthalia Traianou\inst{\ref*{heidelberg},\ref*{planck_bonn}}\orcidlink{0000-0002-1209-6500} \and
  Martin C. Weisskopf\inst{\ref*{nasa_marshall}}\orcidlink{0000-0002-5270-4240} \and
  Kinwah Wu\inst{\ref*{mullard}}\orcidlink{0000-0002-7568-8765} \and
  Francisco Jos\'e Aceituno\inst{\ref*{granada}}\orcidlink{0000-0001-8074-4760} \and
  Giacomo Bonnoli\inst{\ref*{inaf_merate},\ref*{granada}}\orcidlink{0000-0003-2464-9077} \and
  V\'{i}ctor Casanova\inst{\ref*{granada}} \and
  Gabriel Emery\inst{\ref*{granada}} \and
  Juan Escudero\inst{\ref*{harv_smithsonian}, \ref*{granada}}\orcidlink{0000-0002-4131-655X} \and
  Daniel Morcuende\inst{\ref*{granada}}\orcidlink{0000-0001-9400-0922} \and
  Jorge Otero-Santos\inst{\ref*{padova},\ref*{granada}}\orcidlink{0000-0002-4241-5875} \and
  Alfredo Sota\inst{\ref*{granada}}\orcidlink{0000-0002-9404-6952} \and
  Vilppu Piirola\inst{\ref*{physics_finland}}\orcidlink{0000-0003-0186-206X} \and
  George A. Borman\inst{\ref*{crimean}} \and
  Evgenia N. Kopatskaya\inst{\ref*{st.petersb}}\orcidlink{0000-0001-9518-337X} \and
  Elena G. Larionova\inst{\ref*{st.petersb}} \and
  Daria A. Morozova\inst{\ref*{st.petersb}}\orcidlink{0000-0002-9407-7804} \and
  Ekaterina V. Shishkina\inst{\ref*{st.petersb}}\orcidlink{0009-0002-2440-2947} \and
  Sergey S. Savchenko\inst{\ref*{st.petersb},\ref*{pulkovo}}\orcidlink{0000-0003-4147-3851} \and
  Andrey A. Vasilyev\inst{\ref*{st.petersb}}\orcidlink{0000-0002-8293-0214} \and
  Tatiana S. Grishina\inst{\ref*{st.petersb}}\orcidlink{0000-0002-3953-6676} \and
  Ivan S. Troitskiy\inst{\ref*{st.petersb}}\orcidlink{0000-0002-4218-0148} \and
  Alexey V. Zhovtan\inst{\ref*{crimean}} \and
  Callum McCall\inst{\ref*{liverpool}}\orcidlink{0000-0002-3375-3397} \and
  Helen E. Jermak\inst{\ref*{liverpool}}\orcidlink{0000-0002-1197-8501} \and
  Iain A. Steele\inst{\ref*{liverpool}} \and
  Rumen Bachev\inst{\ref*{bulgaria}} \and
  Anton Strigachev\inst{\ref*{bulgaria}} \and
  Ryo Imazawa\inst{\ref*{physics_hiroshima}}\orcidlink{0000-0002-0643-7946} \and
  Mahito Sasada\inst{\ref*{research_tokyo}}\orcidlink{0000-0001-5946-9960} \and
  Yasushi Fukazawa\inst{\ref*{physics_hiroshima},\ref*{astro_hiroshima},\ref*{core_hiroshima}} \and
  Koji S. Kawabata\inst{\ref*{physics_hiroshima},\ref*{astro_hiroshima},\ref*{core_hiroshima}}\orcidlink{0000-0001-6099-9539} \and
  Makoto Uemura\inst{\ref*{physics_hiroshima},\ref*{astro_hiroshima},\ref*{core_hiroshima}}\orcidlink{0000-0002-7375-7405} \and
  Tsunefumi Mizuno\inst{\ref*{astro_hiroshima}}\orcidlink{0000-0001-7263-0296} \and
  Tatsuya Nakaoka\inst{\ref*{astro_hiroshima}} \and
  Sumie Tochihara\inst{\ref*{physics_hiroshima}} \and
  Takahiro Akai\inst{\ref*{physics_hiroshima}} \and
  Hiroshi Akitaya\inst{\ref*{chiba_jap}}\orcidlink{0000-0001-6156-238X} \and
  Andrei V. Berdyugin\inst{\ref*{physics_finland}}\orcidlink{0000-0002-9353-5164} \and
  Masato Kagitani\inst{\ref*{tohoku}} \and
  Vadim Kravtsov\inst{\ref*{physics_finland}}\orcidlink{0000-0002-7502-3173} \and
  Juri Poutanen\inst{\ref*{physics_finland}}\orcidlink{0000-0002-0983-0049} \and
  Takeshi Sakanoi\inst{\ref*{tohoku}}\orcidlink{0000-0002-7146-9020} \and
  Diego \'Alvarez-Ortega\inst{\ref*{forth}, \ref*{physics_crete}}\orcidlink{0000-0002-9998-5238} \and
  Carolina Casadio\inst{\ref*{forth}, \ref*{physics_crete}}\orcidlink{0000-0003-1117-2863} \and
  Sincheol Kang\inst{\ref*{astro_korea}}\orcidlink{0000-0002-0112-4836} \and
  Sang-Sung Lee\inst{\ref*{astro_korea},\ref*{korea}}\orcidlink{0000-0002-6269-594X} \and
  Sanghyun Kim\inst{\ref*{astro_korea},\ref*{korea}}\orcidlink{0000-0001-7556-8504} \and
  Whee Yeon Cheong\inst{\ref*{astro_korea},\ref*{korea}}\orcidlink{0009-0002-1871-5824} \and
  Hyeon-Woo Jeong\inst{\ref*{astro_korea},\ref*{korea}}\orcidlink{0009-0005-7629-8450} \and
  Chanwoo Song\inst{\ref*{astro_korea},\ref*{korea}}\orcidlink{0009-0003-8767-7080} \and
  Shan Li\inst{\ref*{astro_korea},\ref*{korea}} \and
  Myeong-Seok Nam\inst{\ref*{astro_korea},\ref*{korea}}\orcidlink{0009-0001-4748-0211} \and
  Mark Gurwell\inst{\ref*{harv_smithsonian}}\orcidlink{0000-0003-0685-3621} \and
  Garrett Keating\inst{\ref*{harv_smithsonian}}\orcidlink{0000-0002-3490-146X} \and
  Ramprasad Rao\inst{\ref*{harv_smithsonian}}\orcidlink{0000-0002-1407-7944} \and
  Emmanouil Angelakis\inst{\ref*{athens}}\orcidlink{0000-0001-7327-5441} \and
  Alexander Kraus\inst{\ref*{planck_bonn}}\orcidlink{0000-0002-4184-9372} \and
  Petra Benke\inst{\ref*{geo_potsdam}, \ref*{planck_bonn}} \and
  Lena Debbrecht\inst{\ref*{planck_bonn}}\orcidlink{0009-0003-8342-4561} \and
  Julia Eich\inst{\ref*{wurzburg}} \and
  Florian Eppel\inst{\ref*{planck_bonn}, \ref*{wurzburg}}\orcidlink{0000-0001-7112-9942} \and
  Andrea Gokus\inst{\ref*{physics_washington}}\orcidlink{0000-0002-5726-5216} \and
  Steven H\"{a}mmerich\inst{\ref*{remeis_obs}} \and
  Jonas He\ss d\"orfer\inst{\ref*{planck_bonn}, \ref*{wurzburg}}\orcidlink{0009-0009-7841-1065} \and
  Matthias Kadler\inst{\ref*{wurzburg}}\orcidlink{0000-0001-5606-6154} \and
  Dana Kirchner\inst{\ref*{wurzburg}} \and
  Georgios Filippos Paraschos\inst{\ref*{planck_bonn}}\orcidlink{0000-0001-6757-3098} \and
  Florian R\"{o}sch\inst{\ref*{planck_bonn}, \ref*{wurzburg}} \and
  Wladislaw Schulga\inst{\ref*{wurzburg}}
}
\institute{
  Institute of Astrophysics, Foundation for Research and Technology-Hellas, GR-71110 Heraklion, Greece \label{forth} \and
  Department of Physics, University of Crete, 70013, Heraklion, Greece \label{physics_crete} \and
  NASA Marshall Space Flight Center, Huntsville, AL 35812, USA \label{nasa_marshall} \and
  INAF Osservatorio Astronomico di Roma, Via Frascati 33, 00078 Monte Porzio Catone (RM), Italy \label{inaf_obs_rome} \and
  Space Science Data Center, Agenzia Spaziale Italiana, Via del Politecnico snc, 00133 Roma, Italy \label{asi_datacenter} \and
  Center for Astrophysics | Harvard \& Smithsonian, 60 Garden Street, Cambridge, MA 02138 USA \label{harv_smithsonian} \and
  INAF, Istituto di Astrofisica e Planetologia Spaziali, Via Fosso del Cavaliere, 100 - I-00133 Rome, Italy \label{inaf_rome} \and
  ASI - Agenzia Spaziale Italiana, Via del Politecnico snc, 00133 Roma, Italy \label{asi} \and
  Instituto de Astrof\'{i}sica de Andaluc\'{i}a, IAA-CSIC, Glorieta de la Astronom\'{i}a s/n, 18008 Granada, Spain \label{granada} \and
  Max-Planck-Institut für Physik, D-85748 Garching, Germany \label{planck_garching} \and
  Science and Technology Institute, Universities Space Research Association, Huntsville, AL 35805, USA \label{huntsville} \and
  Physics Department and McDonnell Center for the Space Sciences, Washington University in St. Louis, MO, 63130, USA \label{physics_washington} \and
  Université Paris Cité, CNRS, Astroparticule et Cosmologie, F-75013 Paris, France \label{paris} \and
  Institute for Astrophysical Research, Boston University, 725 Commonwealth Avenue, Boston, MA 02215, USA \label{boston} \and
  St. Petersburg State University, 7/9, Universitetskaya nab., 199034 St. Petersburg, Russia \label{st.petersb} \and
  Department of Physics and Astronomy, 20014 University of Turku, Finland \label{physics_finland} \and
  Finnish Centre for Astronomy with ESO, 20014 University of Turku, Finland \label{finnish_eso} \and
  Universit\'{e} de Strasbourg, CNRS, Observatoire Astronomique de Strasbourg, UMR 7550, 67000 Strasbourg, France \label{strasbourg} \and
  Instituto de Radioastronomía Millimétrica, Avenida Divina Pastora, 7, Local 20, E–18012 Granada, Spain \label{mm_granada} \and
  INAF Osservatorio Astronomico di Brera, Via E. Bianchi 46, 23807 Merate (LC), Italy \label{inaf_merate} \and
  Interdisziplin\"{a}res Zentrum f\"{u}r wissenschaftliches Rechnen (IWR), Ruprecht-Karls-Universit\"{a}t Heidelberg, Im Neuenheimer Feld 205, 69120, Heidelberg, Germany \label{heidelberg} \and
  Max-Planck-Institut f\"{u}r Radioastronomie, Auf dem H\"{u}gel 69, D-53121 Bonn, Germany \label{planck_bonn} \and
  Mullard Space Science Laboratory, University College London, Holmbury St Mary, Dorking, Surrey RH5 6NT, UK \label{mullard} \and
  Istituto Nazionale di Fisica Nucleare, Sezione di Padova, 35131 Padova, Italy \label{padova} \and
  Crimean Astrophysical Observatory RAS, P/O Nauchny, 298409, Crimea \label{crimean} \and
  Pulkovo Observatory, St.Petersburg, 196140, Russia \label{pulkovo} \and
  Astrophysics Research Institute, Liverpool John Moores University, Liverpool Science Park IC2, 146 Brownlow Hill, Liverpool, UK \label{liverpool} \and
  Institute of Astronomy and NAO, Bulgarian Academy of Sciences, 1784 Sofia, Bulgaria \label{bulgaria} \and
  Department of Physics, Graduate School of Advanced Science and Engineering, Hiroshima University Kagamiyama, 1-3-1 Higashi-Hiroshima, Hiroshima 739-8526, Japan \label{physics_hiroshima} \and
  Institute of Integrated Research, Institute of Science Tokyo, 2-12-1 Ookayama, Meguro-ku, Tokyo 152-8550, Japan \label{research_tokyo} \and
  Hiroshima Astrophysical Science Center, Hiroshima University 1-3-1 Kagamiyama, Higashi-Hiroshima, Hiroshima 739-8526, Japan \label{astro_hiroshima} \and
  Core Research for Energetic Universe (Core-U), Hiroshima University, 1-3-1 Kagamiyama, Higashi-Hiroshima, Hiroshima 739-8526, Japan \label{core_hiroshima} \and
  Planetary Exploration Research Center, Chiba Institute of Technology 2-17-1 Tsudanuma, Narashino, Chiba 275-0016, Japan \label{chiba_jap} \and
  Graduate School of Sciences, Tohoku University, Aoba-ku,  980-8578 Sendai, Japan \label{tohoku} \and
  Korea Astronomy and Space Science Institute, 776 Daedeok-daero, Yuseong-gu, Daejeon 34055, Korea \label{astro_korea} \and
  University of Science and Technology, Korea, 217 Gajeong-ro, Yuseong-gu, Daejeon 34113, Korea \label{korea} \and
  Section of Astrophysics, Astronomy \& Mechanics, Department of Physics, National and Kapodistrian University of Athens, Panepistimiopolis Zografos 15784, Greece \label{athens} \and
  GFZ Helmholtz Centre for Geosciences, Telegrafenberg, 14476, Potsdam, Germany \label{geo_potsdam} \and
  Julius-Maximilians-Universit\"{a}t W\"{u}rzburg, Institut f\"{u}r Theoretische Physik und Astrophysik, Lehrstuhl f\"{u}r Astronomie, Emil-Fischer-Stra{\ss}e 31, 97074 W\"{u}rzburg, Germany \label{wurzburg} \and
  Dr. Karl-Remeis Sternwarte and Erlangen Centre for Astroparticle Physics, Friedrich-Alexander Universit\"at Erlangen-N\"urnberg, Sternwartstr.~7, 96049 Bamberg, Germany \label{remeis_obs} 
}

  \abstract{
    Polarimetric properties of blazars allow us to put constraints on the acceleration mechanisms that fuel their powerful jets. By studying the multiwavelength polarimetric behaviour of high synchrotron peaked (HSP) and low synchrotron peaked (LSP) blazars, we aim to explore differences in their emission mechanisms and magnetic field structure in the acceleration region. In this study, we take advantage of several X-ray polarisation observations of HSP by the IXPE, including four new observations of Mrk~501, and optical polarisation observations of LSP from RoboPol and many others.
    We find that the polarisation degree (PD) distribution of HSP in X-rays is systematically higher than in optical and mm-radio wavelengths, as reported in previous IXPE publications.
    The distribution of the X-ray electric vector position angles (PA) is centered around the jet axis with most of the observations consistent with zero difference within uncertainties.
    In fact, the distribution of the offset of the PA from the jet axis is consistent between the LSP and HSP populations (with PA measured in optical for the first, X-ray for the latter), suggesting a common magnetic field structure close to the acceleration region, in strong support of the emerging energy stratified picture of particle acceleration followed by energy loss in blazar jets.}

   \maketitle
\nolinenumbers

\section{Introduction}
\label{introduction}

The jets of active galactic nuclei (AGN) are extremely luminous and their emission goes from radio wavelengths up to the extremely energetic $\gamma$-rays.
Their non-thermal radiation is commonly thought to arise from a highly collimated jet of particles accelerated to relativistic energies, extending along the polar axis of an accreting supermassive black hole (SMBH).
In the specific case of blazars, the jet of magnetised plasma is oriented toward our line of sight.
Their spectral energy distribution (SED) shows two humps, commonly known as the synchrotron and the high-energy hump.
Based on the frequency of the synchrotron peak, we classify these sources into three classes: LSP sources (low synchrotron peak), with $\nu_{peak} < 10^{14}$ Hz; ISP sources (intermediate synchrotron peak), with $10^{14}$ Hz $< \nu_{peak} < 10^{15}$ Hz; HSP sources (high synchrotron peak), with $\nu_{peak} > 10^{15}$ Hz.
The acceleration mechanism of these particles is still unclear and under debate.
The synchrotron hump arises from synchrotron radiation emitted by accelerated electrons and positrons and peaks between infrared and X-rays.
The high-energy (HE) component extends from the keV to the TeV energy range, and the responsible emission mechanism is commonly thought to be Compton scattering of photons by higher energy particles \citep{maraschi1992jet}, but the origin of the seed photons is still unclear; some models also propose hadronic processes to contribute to the high-energy flux \citep{boettcher2012modeling}.

There are still many open questions about blazars.
For instance, the particle acceleration mechanism is still under debate.
The primary theoretical models include moving or stationary shocks (i.e., plasma with a turbulent field crossing a shock front, \citealt{marscher1985models}) and magnetic reconnection \citep{sironi2015}.
The magnetic field structure plays a pivotal role in particle acceleration and can be probed through multiwavelength linear polarisation measurements.
In the shock scenario, we expect particles to be accelerated in a small volume and lose energy while being advected away from the shock, encountering increasingly turbulent magnetic field areas, which would bring to a strongly chromatic PD (increasing with frequency); due to the magnetic field of the plasma being parallel to the shock normal, the polarisation angle should be aligned with the jet direction \cite[e.g.,][]{hughes1985}.
In a magnetic reconnection scenario, we would expect lower polarisation degree (hereafter PD) for both optical and X-rays and random polarisation angles (hereafter PA), since the magnetic field is highly disordered \citep{bodo2021kink}.

Optical polarisation observations using RoboPol \citep{robopol} have found that LSP sources in the optical range present stronger polarisation degrees, higher variability and large rotations of the polarisation angle compared to HSP, while HSP polarisation angles in the same range are more stable and have a preferred direction \citep{blinov2016robopol,angelakis2016robopol}. This can be explained by either shock acceleration or magnetic reconnection in a small volume close to the acceleration region: the highest energy particles (emitting at the synchrotron peak and beyond) are expected to be located in this small volume, losing energy as they move away from the acceleration region.
Therefore the highest polarisation degrees are expected close to the synchrotron peak due to the energy stratification, and the higher variability is related to smaller emission volumes. Smooth rotations of the polarisation angle are expected when plasma propagates through regions where the helical component of the magnetic field is dominant, while further along the jet stream the rotations are expected to be stochastic, due to the increase of turbulence in the magnetic field \citep{Marscher2014}. If this is indeed a Universal picture, we expect to find a similar polarisation behaviour of HSP in the X-rays. Notably, the first observations of blazars in X-ray polarisation by the Imaging X-ray Polarimetry Explorer (IXPE, \citealt{weisskopf2022imaging}) have shown a chromatic polarisation degree with a polarisation angle roughly aligned with the jet axis \citep{liodakis2022polarized,di2022x,kouch2024ixpe}, as well as smooth rotations of the polarisation angle  \citep{di2023discovery,maksym2024two}

We have now accumulated a sufficient number of IXPE HSP observations that can be used to more qualitatively probe whether the same characteristics (i.e., strong polarisation degrees and substantial rotations of the polarisation angle) can be found in LSP and HSP in the two different spectral ranges, corresponding to analogous regions of their spectral energy distributions. Here we aim to perform the first X-ray polarisation population study of blazars, and explore the similarities of HSP-X-ray to the LSP-optical polarisation for the first time. In Section \ref{new_obs} we present four new X-ray polarisation observations of Mrk~501 taken during IXPE Cycle 1, and in Section \ref{sources} we present the HSP and LSP samples used in this work.
The results of our population study are presented in Section \ref{results}, and our findings are discussed in Section \ref{discussion}.

\section{New X-ray polarisation observations}
\label{new_obs}

\begin{figure*}[h!]
    \centering
    \includegraphics[width=\textwidth]{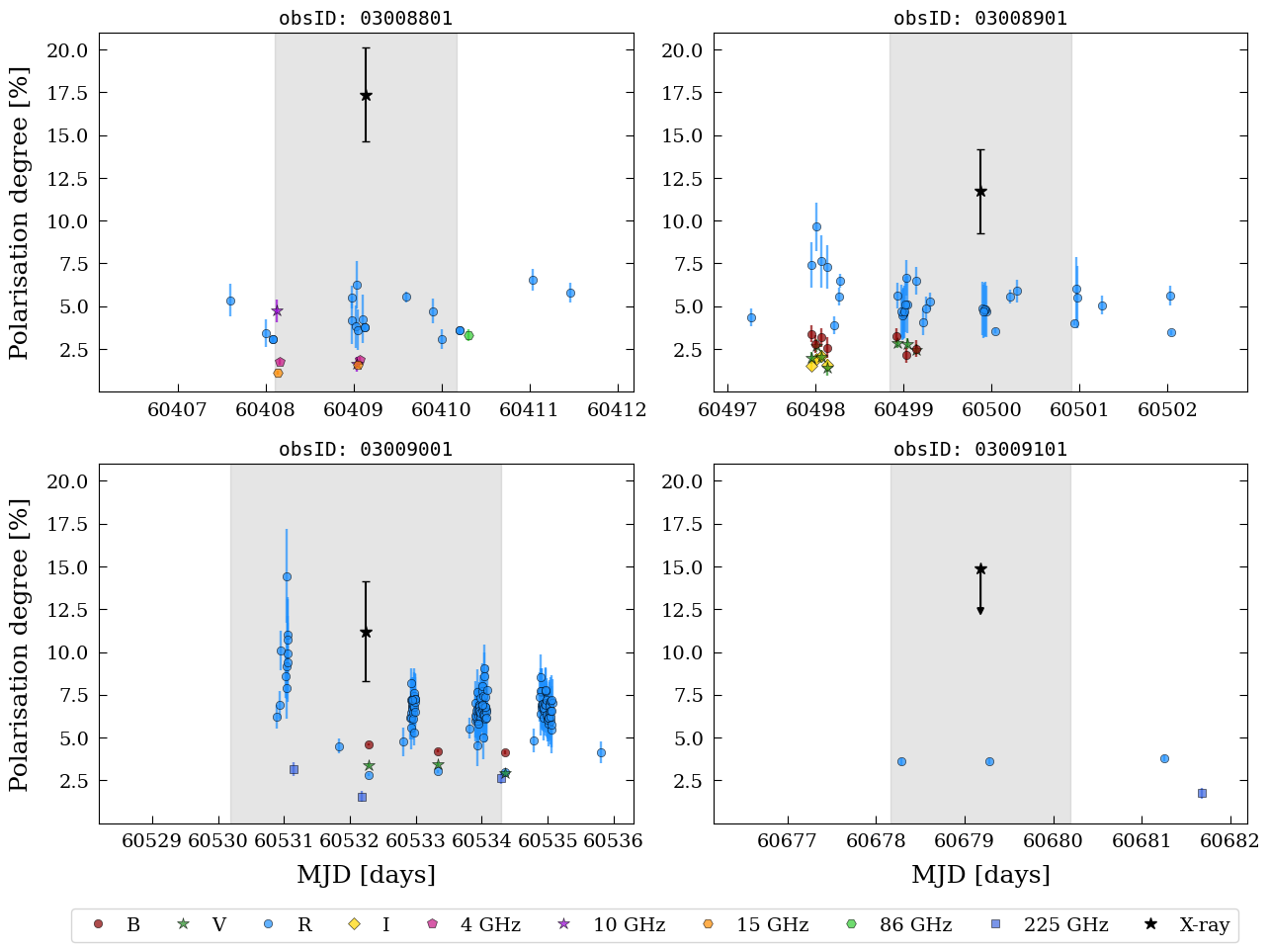}
    \caption{Multiwavelength measurements of the PD of Mrk~501 simultaneous to the new IXPE observations.
    The IXPE observation times are highlighted in grey and its measurements are represented by the black stars.}
    \label{fig:mrk501_pd}
\end{figure*}

\begin{figure*}[h!]
    \centering
    \includegraphics[width=\textwidth]{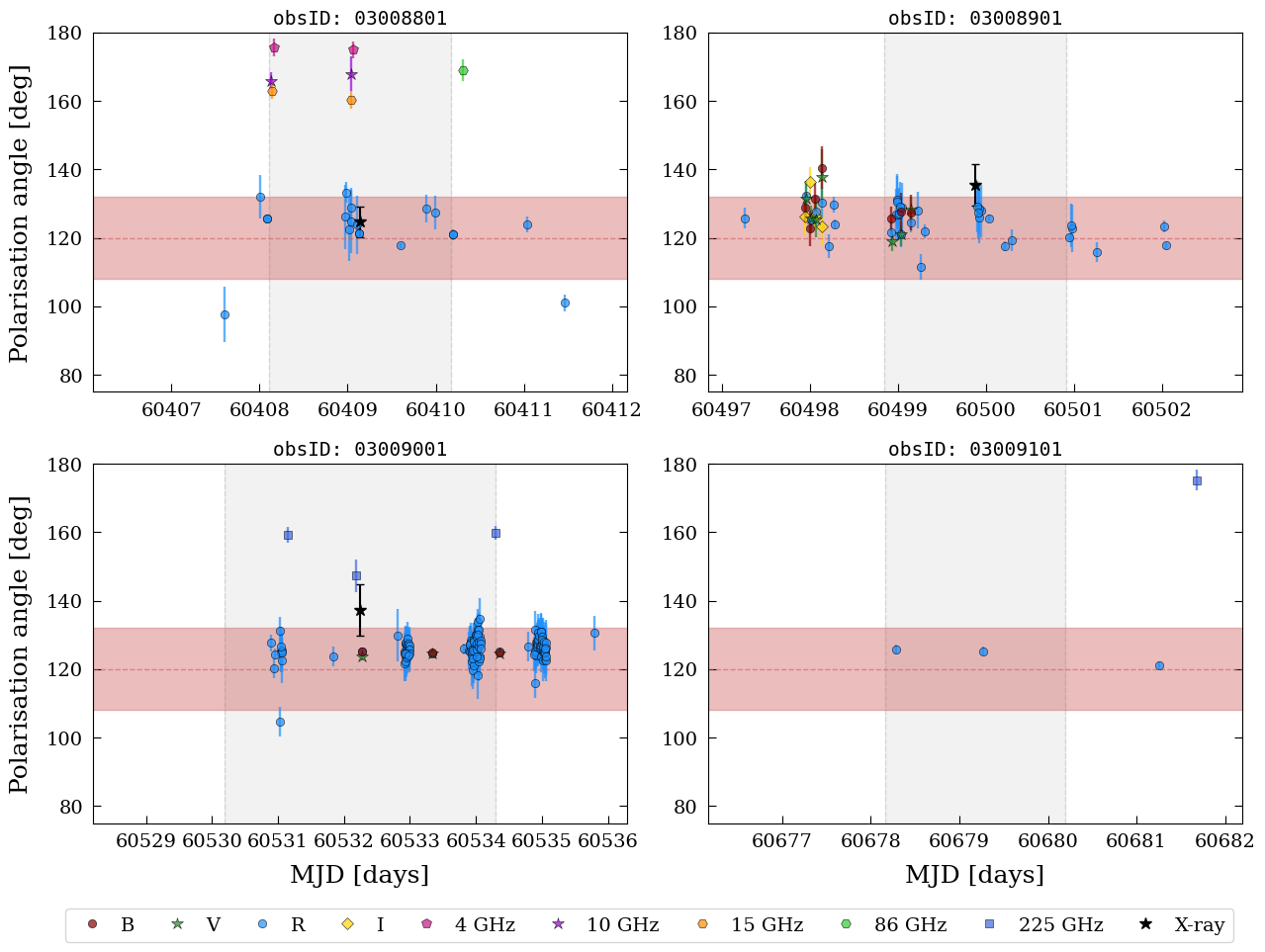}
    \caption{
    Multiwavelength measurements of the PA of Mrk~501 simultaneous to the new IXPE observations.
    The IXPE observation times are highlighted in grey and its measurements are represented by the black stars. The jet direction is represented by the red dashed line, with its uncertainty represented by the red shaded area. The fourth panel is missing the X-ray PA measurement, as the IXPE PD measurement is a $3\sigma$ upper limit and therefore the polarisation angle is not defined.}
    \label{fig:mrk501_pa}
\end{figure*}

IXPE is a joint NASA-ASI mission, launched on December 9, 2021, and it is capable of measuring the degree of polarisation (PD) and the electric vector position angle (PA) in the 2-8 keV energy range.
The analysis of the observations is generally performed using standard spectro-polarimetric analysis \citep{strohmayer2017} through \texttt{XSPEC} \citep{arnaud1996xspec}, with a log-parabola model \citep{massaro2004log} used to fit the X-ray spectra, which represents the typical spectral shape of HSP both in quiescent and flaring states \citep{donnarumma2008june,balokovic2016multiwavelength}.

IXPE observed Mrk~501 four times during 2024 and 2025 (see Table \ref{table:new_obs_mwl} for observation details). 
The first three observations resulted in a significant detection of the polarisation degree, whereas the fourth resulted in an upper limit (Fig. \ref{fig:mrk501_pd}).
For consistency, we follow the analysis as described in previous IXPE-Mrk 501 papers \citep{liodakis2022polarized, chen2024x}.
In line with previous observations, we find a variable polarisation degree between observations, with the polarisation angle aligned with the jet axis in all three cases (Fig. \ref{fig:mrk501_pa}).
We do not detect any polarisation variability within the 100~ksec IXPE exposures following the methodology in \cite{di2023discovery} and \cite{kim2024magnetic}. 

Simultaneously with IXPE observations, several radio and optical telescopes provided multiwavelength polarisation coverage.
The multiwavelength data of the four new Mrk 501 observations are plotted in Fig. \ref{fig:mrk501_pd} (PD) and Fig. \ref{fig:mrk501_pa} (PA); the median of the radio and optical polarisation measurements can be found in Table \ref{table:new_obs_mwl}.
Those are namely the Belogradchik Observatory \citep{bachev2024}, the Calar Alto Observatory \citep{agudo2012, escudero2024iop4}, the IRAM 30m Millimeter Radiotelescope through the POLAMI Program \citep{agudo2018}, the Effelsberg 100m telescope through the TELAMON \citep{Eppel2024} and QUIVER  \citep{kraus2003,myserlis2018} programs, T60 at the Haleakala observatory \citep{Piirola2014,Berdyugin2019,Piirola2021}, the Korean VLBI Network (KVN) array in single-dish mode \citep{kang2015}, the Nordic Optical Telescope \citep{nilsson2018}, the Liverpool Telescope \citep{jermak2018, steele2004, shrestha2020}, the LX-200, the Submillimeter Array through the SMAPOL program (Myserlis et al., in prep.), the Perkins Telescope \citep{jorstadmarscher2016}, the Sierra Nevada Observatory \citep{otero2024}, and the Skinakas observatory \citep{robopol}.
The data analysis procedures and observing strategies are described in detail in \citet{liodakis2022polarized}, \citet{di2023discovery}, \citet{peirson2023x}, \citet{kouch2024ixpe}, \citet{kouch2024ixpeflare}.
Polarisation measurements in the R band have been corrected for the host-galaxy contribution following \citet{nilsson2007host} and \citet{hovatta2016optical}.
Mrk~501 has shown consistent behaviour in all the X-ray polarisation observations so far, where all X-ray-optical-radio PA are aligned with the jet axis measured at 43 GHz, and the X-ray PD is typically $>2-3$ times higher than the optical PD and typically $>5-6$ higher than the radio PD.

\section{Samples}
\label{sources}

This section aims to present the chosen sample for the LSP and HSP populations (ten sources and six sources, respectively).
A detailed description of the sources taken into account and the observations used for this analysis can be found in Appendix \ref{appendix_sample}, Table \ref{table:sample}.
Specifically, a more detailed description of the HSP sample and the previous studies published by the IXPE collaboration can be found in Table \ref{table:hsp_obs}.

\subsection{HSP sample}
\label{hsp_sources}

For the HSP sample, we use previously published multiwavelength polarisation data and jet direction of blazars (see Table \ref{table:hsp_obs} for details), taken from previous works by the IXPE collaboration.
Our sample consists of the following sources: 1ES~0229$+$200 \citep{ehlert2023x}, Mrk~421 \citep{di2022x, di2023discovery, kim2024magnetic, maksym2024two}, PG~1553$+$113 \citep{middei2023ixpe}, Mrk~501 \citep{chen2024x, liodakis2022polarized},  1ES~1959$+$650 \citep{errando2024detection, pacciani2025}, PKS~2155$-$304 \citep{kouch2024ixpe}, in order of increasing RA.

We only consider strictly simultaneous observations in the optical and mm-radio bands.
For some observations of Mrk~421 and 1ES~1959+650 we use time-resolved results of the polarisation analysis, instead of the average values (this was done for observations from \citealt{di2023discovery} and \citealt{maksym2024two}, and for the second observation from \citealt{errando2024detection}).
In particular, for 1ES~1959+650 we only use one of the four time bins reported by the authors, as the other three are upper limits.
These sources are the only two in the sample presenting large deviations of the PA from the jet angle.

We filter data by selecting only $3 \sigma$ detections ($\rm PD_{value}/PD_{error} > 3$).
For the optical observations, we consider the R band measurements since it is the only optical band corrected for the host galaxy contribution, and for the mm-radio observations the highest available frequency, typically 225 GHz.

\subsection{LSP sample}
\label{lsp_sources}

We start our sample selection from the RoboPol monitoring program \citep{blinov2021}, which includes observations of 222 AGN at Dec. $>-25^\circ$ from 2013 to 2017 in the R band; however, only a small subset of 61 sources was monitored regularly.
We select all LSP sources \citep{ajello2020fourth} with available jet directions from \citet{weaver2022kinematics}.
Our final sample consists of the common ten sources, namely PKS~0420$-$01, 4C~71.07, PKS~1510$-$089, 4C~38.41, 3C~345, 4C~09.57, BL Lac, 3C~446, 4C~11.69, 3C~454.3, in order of increasing RA.
For 3C~454.3 and BL Lac we include additional optical polarisation observations from the IXPE campaigns (\citealp{middei2022x,peirson2023x,marshall2024observations, agudo2025}).

The initial sample also included 3C 273; however, this source was removed from this study, as its strong accretion disk contribution to the optical emission could lower the overall polarisation degree.

\section{Population analysis}
\label{results}

\subsection{X-ray polarisation analysis}
\label{xray_pol}

\begin{figure}[h!]
    \centering
    \includegraphics[width=1\linewidth]{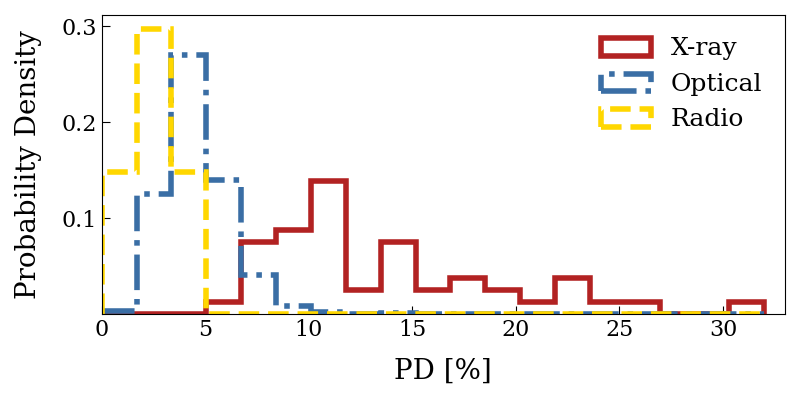}
    \caption{Measured PD of all HSP observations in each spectral range: X-ray (red solid line), optical (blue dash-dotted line), mm-radio (yellow dashed line).
    The histogram is normalised due to the low number of radio observations with respect to the other bands.}
    \label{fig:pd_hsp}
\end{figure}

\begin{figure}[h!]
    \centering
    \includegraphics[width=1\linewidth]{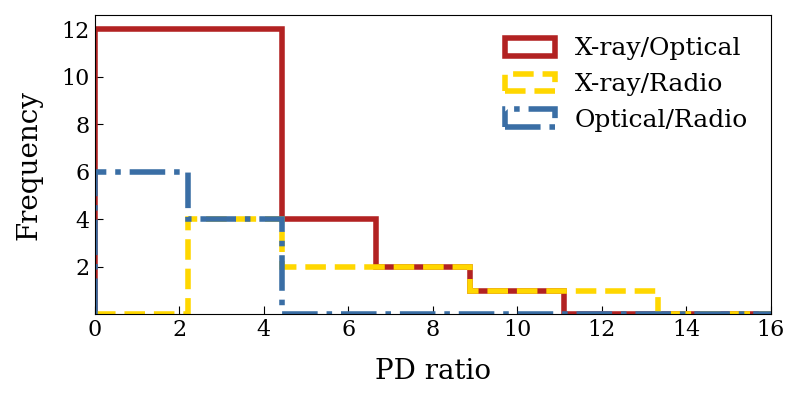}
    \caption{Ratio of HSP PD between the different ranges, computed for all HSP observations: X-ray over optical (red solid line), X-ray over mm-radio (yellow dashed line), optical over mm-radio (blue dash-dotted line).}
    \label{fig:pd_ratio_hsp}
\end{figure}

Using the simultaneous multiwavelength HSP observations we plot the distribution of the X-ray, optical and radio polarisation degree (Fig. \ref{fig:pd_hsp}) and their ratios (Fig. \ref{fig:pd_ratio_hsp}).
Note that these figures represent all of the observations taken into account for the six HSP sources in our sample: separate observations of the same source sample the same underlying magnetic field structure and particle acceleration properties of that object.
The different observations are shown for completeness, but the median values of the polarisation properties as likely more representative.
The median values of PD and PA for each source can be found in Table \ref{table:sample}.
As noted in previous publications discussing IXPE observations, the X-ray PD is consistently higher than the optical, which in turn is higher than the radio, in line with the energy stratification of the emission owing to particle cooling.
The median X-ray to optical ratio of our HSP sample is 2.5, with standard deviation 2.0; the median X-ray to radio ratio is 6.0, with standard deviation  2.5; the median Optical to radio ratio is 1.7, with standard deviation 0.7.
Note that radio measurements were not available for every IXPE observation interval, so the number of radio observations included in the analysis is lower than the number of optical and X-ray observations.

\begin{figure}[h!]
    \centering
    \includegraphics[width=1\linewidth]{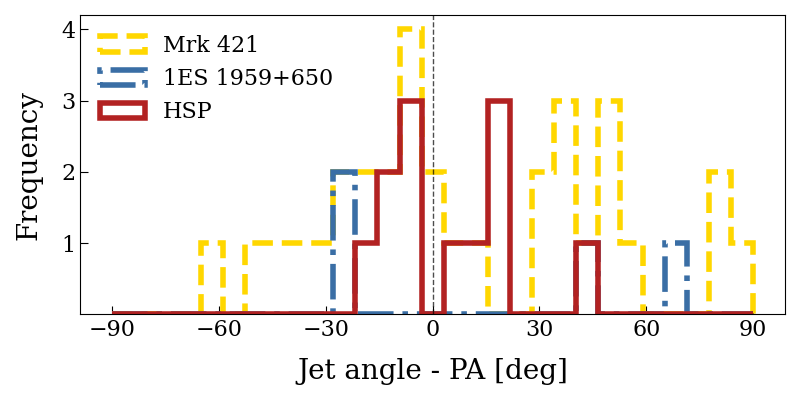}
    \caption{Histogram of the angular difference between the jet and the X-ray PA, computed for every HSP observation.
    Mrk~421 (yellow dashed line) and 1ES~1959+650 (blue dash-dotted line) are plotted separately from the others, as they present large amplitude rotations of the PA during the IXPE observation periods.
    The other sources are plotted altogether (red solid line).}
    \label{fig:hist_adiff_hsp}
\end{figure}

We also explore the difference between the measured PA and the jet direction in the X-ray range for every observation.
This is particularly important because different acceleration mechanisms correspond to different angles between the magnetic field and the jet axis.
The results are shown in Figure \ref{fig:hist_adiff_hsp}.
The X-ray PA offset from the jet axis is centered around zero with a median of 2.5° and a standard deviation of 17.5° (due to the two sources that present large departures of the PA from the jet).
This would suggest that during quiescent or average states, the X-ray PA in HSP fluctuates about the jet axis. Large deviations (>50°) from the jet axis alignment are observed only for Mrk~421, during rotations of the polarisation angle, and for 1ES~1959+650 (plotted separately in Fig. \ref{fig:hist_adiff_hsp}).
Note that this figure also reports all observations separately, to underline the wide range of PA values measured for each source.
This is similar to the optical polarisation behaviour observed in LSP during PA rotations, and it is often attributed to the emergence of new polarisation components crossing the radio core \citep{liodakis2020two,kouch2024ixpeflare}.

\subsection{HSP-X-ray compared to LSP-optical}
\label{hsp_lsp_results}

\begin{figure}[h!]
    \centering
    
    \begin{subfigure}{0.9\linewidth}
        \centering
        \includegraphics[width=\linewidth]{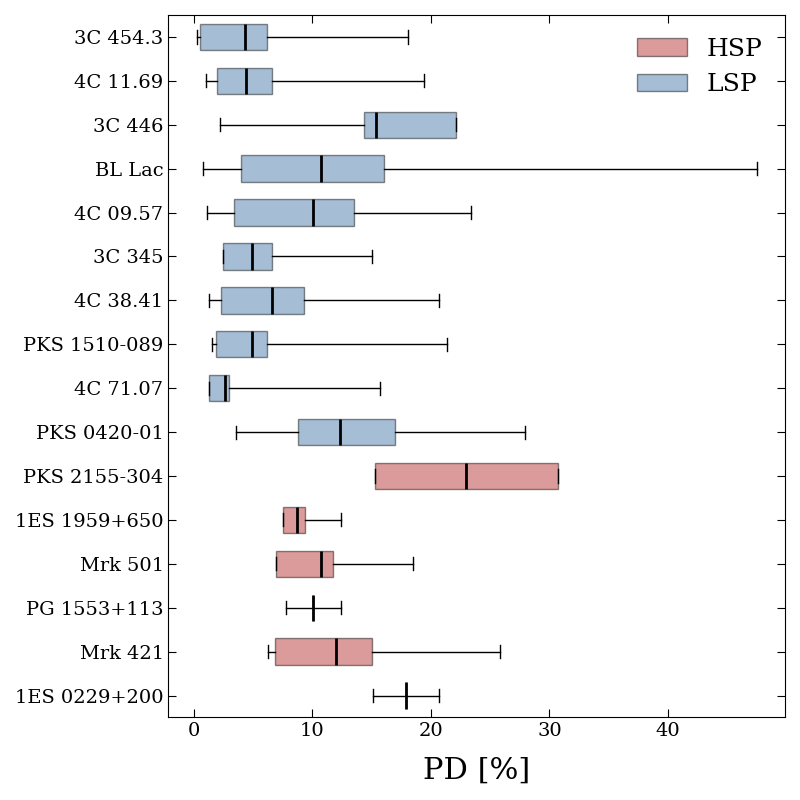}
        \caption{}
        \label{fig:boxplot_pd}
    \end{subfigure}
    
    \vspace{0.2cm}
    
    \begin{subfigure}{0.9\linewidth}
        \centering
        \includegraphics[width=\linewidth]{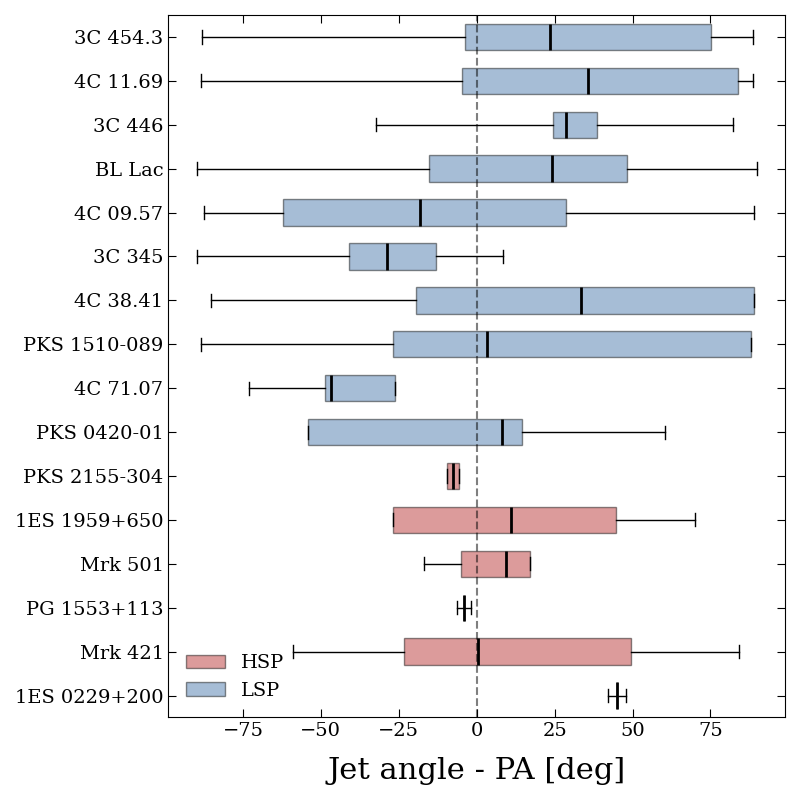}
        \caption{}
        \label{fig:boxplot_adiff}
    \end{subfigure}

    \caption{
    The first boxplot represents the PD of the sources in our sample (LSP in blue, above; HSP in red, below).
    The second boxplot represents the angular difference between the jet direction and the PA of the sources, with the dashed black line indicating a difference of 0° (LSP in blue, above; HSP in red, below).
    The black vertical line inside the coloured box marks the median over all observations.
    The coloured box represents the shortest interval containing 68\% of the observations, while the horizontal black lines extend to the full range of values measured across all observations considered.
    Note that for PKS~2155$-$304 the coloured box covers the full range, as we only had two observations; for PG~1553$+$113 and 1ES~0229$+$200 we only had one observation, so we represent here the measurement and its error.
    }
    \label{fig:boxplot_combined}
\end{figure}

We then proceed with a comparison between the HSP sample and the LSP sample.
The analysis focuses on two different wavelength ranges (X-rays for HSP and optical for LSP), as these correspond to analogous regions of their spectral energy distributions.
The PD distributions of the two samples are shown in Figure \ref{fig:boxplot_pd}.
The median HSP PD is 11.4\%, with standard deviation 5.1\%, while for LSP it is 5.8\%, with standard deviation 4.0\%.
The HSP generally show higher PD than the LSP (confirmed by the Wilcox test applied on the two distributions); however, this may be attributed to an observational bias at X-ray wavelengths: the combination of the typical IXPE exposure of 100 ksec and the typical HSP flux ($\approx 10^{-10} erg \, s^{-1} cm^{-2}$) yields a minimum detectable polarisation degree at the 99\% confidence interval ($\rm MDP_{99}$, \citealt{weisskopf2010understanding}) of about 5-6\%.
Since lower values cannot be measured (and indeed we have omitted upper limits in Mrk~501 and 1ES~1959+650), the X-ray PD distribution is skewed to higher values.
The disk emission can also lower the optical PD of LSP, as it is unpolarised and strong in the optical range (but not in the X-rays, and therefore not for HSP).

\begin{figure}[h!]
    \centering
    \includegraphics[width=0.9\linewidth]{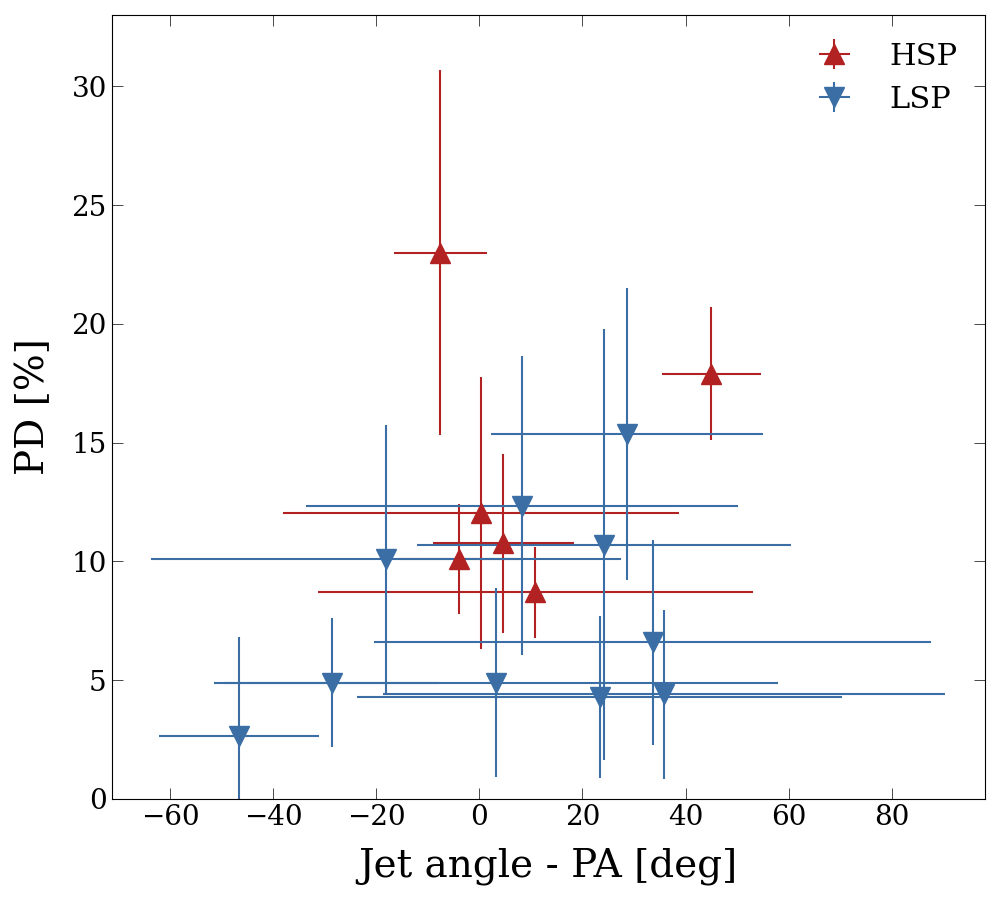}
    \caption{
    The median PA offset from the jet plotted versus the median PD of each object (in X-rays for HSP, red upward triangles; in optical for LSP, blue downward triangles).
    Note that the median values are the most crucial for this study, as they represent the average behaviour of an object.
    }
    \label{fig:pd_vs_adiff}
\end{figure}

We also repeat the analysis of the PA offset from the jet described in the previous section for the LSP sample. 
Both HSP and LSP distributions are shown in Figs. \ref{fig:boxplot_adiff} and \ref{fig:pd_vs_adiff}.
We find similar results for the LSP, with a median of 15.87\% and a standard deviation of 27.1\%.
To compare the distributions of PD and of PA alignment of the two samples, we apply the Anderson-Darling test (hereafter A-D).
The A-D is a non-parametric test to assess whether the observations come from the same parent population \citep{andersondarling}.
We set a p-value threshold to reject the null hypothesis, stating that the two distributions come from the same parent distribution, at 0.05.
We apply the test on the distributions of median PD and of median PA offset from the jet direction.
The test results are reported in Table \ref{tests_table}.
We also compare the LSP sample to the HSP sample excluding Mrk~421, to assess whether the much larger number of Mrk~421 observatons (compared to other HSP) has an effect on our comparison.
Considering the whole HSP sample, we find consistency in the PA offset distributions and mild inconsistency between the PD distributions. Excluding Mrk~421 does not alter the results for the PA offset, but yields a result of consistency for the PD distributions.
Our results on the PA analysis point towards the possibility of a similar magnetic field configuration in the acceleration region for HSP and LSP.
The results of these comparisons will be discussed in detail in the next section.

\begin{figure}[h!]
    \centering
    \includegraphics[width=1\linewidth]{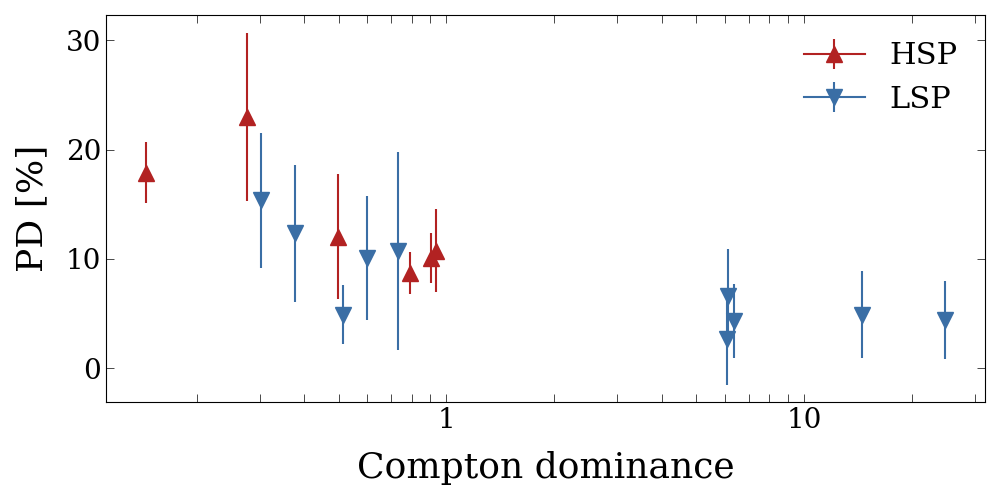}
    \caption{The Compton dominance of the sources (logarithmic scale) plotted against their median PD (in the X-rays for HSP, red upward triangles, and optical for LSP, blue downward triangles).
    }
    \label{fig:compdom_vs_pd}
\end{figure}

Finally, we search for similarities in the behaviour of the Compton dominance of LSP and HSP.
This quantity is defined as the ratio between the energy flux per logarithmic interval of frequency $\nu F_{\nu, HE}$ at the high-energy peak frequency and $\nu F_{\nu, syn}$ at the synchrotron peak frequency and could indicate the dominant process in the high-energy emission.
HSP sources typically have Compton dominance values lower than one, while LSP can greatly exceed unity.
We use the values listed in the Fermi 4LAC catalog \citep{ajello2020fourth} to compute the Compton dominance for each source and plot it against the median X-ray PD in the case of HSP and the median optical PD for LSP (Fig. \ref{fig:compdom_vs_pd}).
Note that the Compton dominance values are calculated from the synchrotron and high-energy peak fluxes averaged over the decades of observations.
Nevertheless, our results highlight an anticorrelation between the Compton dominance and the polarisation degree, and a fairly good consistency between the HSP and LSP samples.

\begin{table}[h!]
\centering
\caption{
P-values resulting from the A-D test applied to the HSP and LSP median PD distributions, and to the distributions of the median difference PA-jet direction.}
\begin{subtable}[t]{0.45\textwidth}
    \centering
    \begin{tabular}{c c}      
        \hline\hline
        \noalign{\vskip 0.4mm}
        Distribution & p-value \\  
        \hline
        \noalign{\vskip 0.6mm}
        PD & 0.043 \\
        \noalign{\vskip 0.2mm}
        \hline
        \noalign{\vskip 0.6mm}
        PA offset & 0.250 \\
        \noalign{\vskip 0.2mm}
        \hline
    \end{tabular}
    \caption{\centering LSP sample compared to the whole HSP sample.}
    \label{subtable_all}
\end{subtable}
\vskip 4mm
\begin{subtable}[t]{0.45\textwidth}
    \centering
    \begin{tabular}{c c}      
        \hline\hline
        \noalign{\vskip 0.6mm}
        Distribution & p-value \\  
        \hline
        \noalign{\vskip 0.6mm}
        PD & 0.056 \\
        \noalign{\vskip 0.2mm}
        \hline
        \noalign{\vskip 0.6mm}
        PA offset & 0.250 \\
        \noalign{\vskip 0.2mm}
        \hline
    \end{tabular}
    \caption{\centering LSP sample compared to HSP sample excluding Mrk~421.}
    \label{subtable_without421}
\end{subtable}
\label{tests_table}
\end{table}

\section{Discussion}
\label{discussion}

Figure \ref{fig:pd_hsp}, representing all observations taken into account for this work, highlights that the mm-radio PD is generally lower than the optical PD, which in turn is generally lower than the X-ray PD.
In fact, a similar behaviour had previously been observed in BL Lacertae (LSP/ISP source) during a flaring state \citep{peirson2023bllac} in which the high-energy tail of the synchrotron emission was detected by IXPE.
Similarly to HSP, BL Lacertae showed a higher X-ray PD than optical and mm-radio, with the PA aligned with the jet axis within uncertainties.
The overall observed multiwavelength polarisation behaviour is naturally explained by shock acceleration, which starts with a well-ordered magnetic field within the acceleration region, growing progressively more turbulent further downstream, resulting in a decreasing PD with longer wavelengths \citep{liodakis2022polarized}.
However, the dominant underlying process driving the decrease in PD is particle cooling and advection from the acceleration region, which can also occur under alternative acceleration scenarios.
In absence of alternative scenario simulations demonstrating a chromatic PD, we still favour shocks as the dominant acceleration mechanism, as also supported by shock acceleration simulations \citep{tavecchio2018, sciaccaluga2025}.

Recently, an alternative model has been proposed by \citealt{bolis2024}, which find a similar solution through confined, magnetically dominated jets, where the polarisation strongly depends on the magnetic field configuration.
Under this scenario, our observations would suggest that the jet is nearly parabolic in the acceleration region. This is in fact supported by VLBI observations of blazars and nearby radio galaxies \citep{kovalev2020}.

As for the comparison between HSP and LSP, we do find mild inconsistencies between the PD distributions of the two populations, which appear not to be drawn from the same parent population according to the resulting p-value.

The offset between the PA and the jet direction is centered around zero for both HSP (median 2.5°, standard deviation 17.5°) and LSP (median 15.8°, standard deviation 27.1°), with the A-D test suggesting that we cannot reject the hypothesis that both distributions come from the same parent population.
Over a total number of 2247 LSP observations, 876 ($\approx 39\%$) present a PA offset consistent with 0 within $3 \sigma$; for HSP, over a total of 48 data points, 30 ($\approx 63\%$) do.
For each comparison, we apply the test on the distribution of median values, as the median represents the average behaviour of the source.
Our results suggest that the magnetic field configuration in the acceleration region is the same for HSP and LSP, providing strong evidence for the energy-stratified picture.

Previous studies have shown that during flaring states, the variability of the polarisation properties can be associated with moving components within the jet, which result in rotations of the PA \citep{marscher2008,marscher2010}.
There are several notable examples in optical polarisation for LSP \citep{liodakis2020two,kouch2024ixpeflare} and possibly in X-ray polarisation for HSP \citep{kim2024magnetic} in further support of this picture.

\citet{blinov2016robopol} demonstrated that not all blazars show rotations of the optical PA, but rather a smaller sample that is predominantly LSP.
Specifically, only 13 out of 33 LSP (39\%) in the monitored RoboPol sample showed rotations during the three-year program.
Of the six HSP observed by IXPE, only Mrk~421 and 1ES~1959+650 show large rotations (33\%).
Given the similarity of the LSP-optical to the HSP-X-ray polarisation found in this work, it is then not unreasonable to expect only two sources to show rotations.
More IXPE observations of alternative HSP sources will provide the opportunity to discover more rotations of the PA and constrain their origin.
Interestingly, PG~1553+113 has shown optical rotations \citep{blinov2016robopol,blinov2018,middei2023ixpe}, making it a prime candidate for detecting X-ray PA rotations.
The rotations observed in the optical range for the LSP were often associated to $\gamma$-ray flares \citep{blinov2016,blinov2018}.
It is then reasonable to expect that the observed X-ray PA rotations in HSP may be associated to VHE $\gamma$-ray flares, as previously observed in Mrk 421 \citep{magic2025}.
This highlights the promising synergy between IXPE and upcoming TeV facilities like CTA \citep{cta2022}.

For both HSP and LSP, we find a general decrease of polarisation degree with increasing Compton dominance.
Of the five LSP sources presenting Compton dominance lower than 1, two are classified as BL Lac objects, and three as flat-spectrum radio quasars \citep{ajello2020fourth}.
The strong anticorrelation  between PD and Compton dominance is confirmed by the Spearman test, which yields a p-value of 0.00004 and a correlation coefficient $r$ of $-0.84$.
This anticorrelation can be interpreted as due to the increasing contribution of the high-energy component of the SED to the observed bands. The observed high-energy side of the synchrotron spectrum is in reality the superposition of both synchrotron and high-energy hump emission: while the synchrotron emission is polarised, the low-energy side of the high-energy emission component is likely dominated by the less polarised Compton scattering emission \citep{agudo2025, liodakis2025}.
As the contribution of the Compton emission becomes dominant over the total emission, the overall PD is lowered.
To confirm or discard this hypothesis, a more thorough study of this relation considering simultaneous data should be carried out.

\section{Conclusions}
\label{conclusions}

In this work we compare two blazar populations consisting of LSP and HSP, looking for similarities in their polarisation properties, in particular in the PD and the difference between the jet direction and the PA.

The main results of this study can be summarised as follows:

\begin{itemize}
    \item Four new IXPE observations of Mrk~501 support previous results of the X-ray PD generally being $>2-3 \, \times$ larger than the optical and $>5-6 \, \times$ larger than the millimetre radio.
    \item The HSP population shows chromaticity, where the PD increases with increasing frequency (Figure \ref{fig:pd_hsp}).
    The median of the X-ray to optical and X-ray to radio PD ratios are 2.5 and 6.0, respectively. This supports the energy stratified picture where particles are accelerated and then advected from the acceleration region as they cool down \citep{liodakis2022polarized}.
    \item The X-ray PA are typically aligned with the jet axis, except during large rotations of the polarisation angle (as observed in Mrk~421).
    Combined with the chromaticity of the PD, this provides further evidence for shock acceleration in blazar jets.
    \item The optical PA of LSP and the X-ray PA of HSP show consistent behaviour, possibly suggesting a common energy-stratification scenario for blazar jets, as predicted in \citet{angelakis2016robopol}.
    The detection of X-ray PA rotations in Mrk~421 and 1ES~1959+650 further confirms this hypothesis, suggesting that some HSP can show X-ray PA rotations, although not all HSP will, similarly to previous results for LSP rotations in the optical range \citep{blinov2016robopol}.
    Observing more HSP could prove beneficial in detecting additional X-ray rotations, with PG~1553+113 likely being a prime candidate.
    \item The PD of both LSP and HSP decreases with increasing Compton dominance.
    This is consistent with the hypothesis of a superposition of synchrotron and Compton emission, with the first being polarised and the latter unpolarised. A more thorough study of this relation should be performed by considering the simultaneous SED, rather than the long-term averages considered here.

\end{itemize}

\begin{acknowledgements}

      This research has made use of data from the RoboPol programme, a collaboration between Caltech, the University of Crete, IA-FORTH, IUCAA, the MPIfR, and the Nicolaus Copernicus University, which was conducted at Skinakas Observatory in Crete, Greece.

      The Imaging X-ray Polarimetry Explorer (IXPE) is a joint US and Italian mission.  The US contribution is supported by the National Aeronautics and Space Administration (NASA) and led and managed by its Marshall Space Flight Center (MSFC), with industry partner Ball Aerospace (now, BAE Systems).  The Italian contribution is supported by the Italian Space Agency (Agenzia Spaziale Italiana, ASI) through contract ASI-OHBI-2022-13-I.0, agreements ASI-INAF-2022-19-HH.0 and ASI-INFN-2017.13-H0, and its Space Science Data Center (SSDC) with agreements ASI-INAF-2022-14-HH.0 and ASI-INFN 2021-43-HH.0, and by the Istituto Nazionale di Astrofisica (INAF) and the Istituto Nazionale di Fisica Nucleare (INFN) in Italy.  This research used data products provided by the {\it IXPE} Team (MSFC, SSDC, INAF, and INFN) and distributed with additional software tools by the High-Energy Astrophysics Science Archive Research Center (HEASARC), at NASA Goddard Space Flight Center (GSFC).
      
      The IAA-CSIC co-authors acknowledge financial support from the Spanish "Ministerio de Ciencia e Innovaci\'{o}n" (MCIN/AEI/ 10.13039/501100011033) through the Center of Excellence Severo Ochoa award for the Instituto de Astrof\'{i}sica de Andaluc\'{i}a-CSIC (CEX2021-001131-S), and through grants PID2019-107847RB-C44 and PID2022-139117NB-C44. Some of the data are based on observations collected at the Observatorio de Sierra Nevada; which is owned and operated by the Instituto de Astrof\'isica de Andaluc\'ia (IAA-CSIC), and at the Centro Astron\'{o}mico Hispano en Andaluc\'ia (CAHA); which is operated jointly by Junta de Andaluc\'{i}a and Consejo Superior de Investigaciones Cient\'{i}ficas (IAA-CSIC). The POLAMI observations reported here were carried out at the IRAM 30m Telescope. IRAM is supported by INSU/CNRS (France), MPG (Germany) and IGN (Spain). The Submillimeter Array is a joint project between the Smithsonian Astrophysical Observatory and the Academia Sinica Institute of Astronomy and Astrophysics and is funded by the Smithsonian Institution and the Academia Sinica. Maunakea, the location of the SMA, is a culturally important site for the indigenous Hawaiian people; we are privileged to study the cosmos from its summit. E.L. was supported by Academy of Finland projects 317636 and 320045. 
      
      The research at Boston University was supported in part by National Science Foundation grant AST-2108622, NASA Fermi Guest Investigator grant 80NSSC23K1507, NASA {\it NuSTAR} Guest Investigator grant 80NSSC24K0565, and NASA Swift Guest Investigator grant 80NSSC23K1145. The Perkins Telescope Observatory, located in Flagstaff, AZ, USA, is owned and operated by Boston University. This work was supported by NSF grant AST-2109127. 
      
      This work was supported by JST, the establishment of university fellowships towards the creation of science technology innovation, Grant Number JPMJFS2129. This work was supported by Japan Society for the Promotion of Science (JSPS) KAKENHI Grant Numbers JP21H01137. This work was also partially supported by Optical and Near-Infrared Astronomy Inter-University Cooperation Program from the Ministry of Education, Culture, Sports, Science and Technology (MEXT) of Japan. We are grateful to the observation and operating members of Kanata Telescope.
      
      S. Kang, S.-S. Lee, W. Y. Cheong, S.-H. Kim, H.-W. Jeong, C. Song, S. Li, and M.-S. Nam were supported by the National Research Foundation of Korea (NRF) grant funded by the Korea government (MIST) (2020R1A2C2009003, RS-2025-00562700).
      The KVN is a facility operated by the Korea Astronomy and Space Science Institute. The KVN operations are supported by KREONET (Korea Research Environment Open NETwork) which is managed and operated by KISTI (Korea Institute of Science and Technology Information). 
      
      S. C., B. A-G., and I. L. were funded by the European Union ERC-2022-STG - BOOTES - 101076343. Views and opinions expressed are however those of the author(s) only and do not necessarily reflect those of the European Union or the European Research Council Executive Agency. Neither the European Union nor the granting authority can be held responsible for them.
      
      The data in this study include observations made with the Nordic Optical Telescope, owned in collaboration by the University of Turku and Aarhus University, and operated jointly by Aarhus University, the University of Turku and the University of Oslo, representing Denmark, Finland and Norway, the University of Iceland and Stockholm University at the Observatorio del Roque de los Muchachos, La Palma, Spain, of the Instituto de Astrofisica de Canarias. The data presented here were obtained in part with ALFOSC, which is provided by the Instituto de Astrof\'{\i}sica de Andaluc\'{\i}a (IAA) under a joint agreement with the University of Copenhagen and NOT. We acknowledge funding to support our NOT observations from the Finnish Centre for Astronomy with ESO (FINCA), University of Turku, Finland (Academy of Finland grant nr 306531).
      
      This research was partially supported by the Bulgarian National Science Fund of the Ministry of Education and Science under grants KP-06-H68/4 (2022), KP-06-H78/5 (2023) and KP-06-H88/4 (2024). 
      
      The Liverpool Telescope is operated on the island of La Palma by Liverpool John Moores University in the Spanish Observatorio del Roque de los Muchachos of the Instituto de Astrofisica de Canarias with financial support from the UKRI Science and Technology Facilities Council (STFC) (ST/T00147X/1).

      Partly based on observations with the 100-m telescope of the MPIfR (Max-Planck-Institut f\"ur Radioastronomie) at Effelsberg. Observations with the 100-m radio telescope at Effelsberg have received funding from the European Union’s Horizon 2020 research and innovation programme under grant agreement No 101004719 (ORP).
      F.E., S.H., J.H., M.K., and F.R. acknowledge support from the Deutsche Forschungsgemeinschaft (DFG, grants 447572188, 434448349, 465409577).

      G. F. P. acknowledges support by the European Research Council advanced grant “M2FINDERS – Mapping Magnetic Fields with INterferometry Down to Event hoRizon Scales” (Grant No. 101018682).

      C.C., D.B. and D.A. acknowledge support from the European Research Council (ERC) under the Horizon ERC Grants 2021 programme under grant agreement No. 101040021.

      P.K. was supported by Academy of Finland projects 346071 and 345899. P.K. acknowledges support from the Mets\"ahovi Radio Observatory of Aalto University.

      J.O.-S. acknowledges founding from the Istituto Nazionale di Fisica Nucleare Cap. U.1.01.01.01.009.

\end{acknowledgements}

\newpage

\bibliographystyle{aa} 
\bibliography{bibliography.bib}

\newpage

\begin{appendix}
\onecolumn

\section{Sample and observations}
\label{appendix_sample}

\begin{table*}[h]
\caption{Detailed description of the sample of HSP sources (first 6 sources) and the LSP sources (last 10 sources).}
\centering
\begin{tabular}{c c c c c c}
\hline\hline
\noalign{\vskip 0.3mm}
Source & Time span of observations & Number of observations & Jet direction [deg] & PD [\%] & PA [deg] \\
\noalign{\vskip 0.2mm}
\hline
\noalign{\vskip 0.4mm}
    1ES~0229+200 & Jan 15 - Feb 1, 2023 & 1 & 70.0 $\pm$ 8.5 & 17.9 $\pm$ 2.8 & 25.0 $\pm$ 4.6 \\
    Mrk~421 & May 4, 2022 - Dec 22, 2023 & 4 & Variable & 12.0 $\pm$ 5.7 & -2.14 $\pm$ 31.5 \\
    PG~1553+113 & Feb 1 - 9, 2023 & 1 & 90.0 $\pm$ 8.5 & 10.1 $\pm$ 2.3 & 86.0 $\pm$ 8.0 \\
    Mrk~501 & Mar 8, 2022 - Jan 5, 2025 & 8 & -60.0 $\pm$ 12.0 & 10.8 $\pm$ 3.8 & -50.7 $\pm$ 12.6  \\
    1ES~1959+650 & May 3, 2022 - Aug 19, 2023 & 4 & Variable & 8.7 $\pm$ 1.9 & -16.65 $\pm$ 47.2 \\
    PKS~2155-304 & Oct 27 - Nov 7, 2023 & 1 & -45.0 $\pm$ 8.5 & 23.0 $\pm$ 7.7 & -52.6 $\pm$ 2.9 \\
\hline
\noalign{\vskip 0.4mm}
    PKS~0420-01 & Oct 28,2013 - Nov 4, 2015 & 10 & -61.8 $\pm$ 19.4 & 12.4 $\pm$ 6.3 & -8.55 $\pm$ 52.3 \\
    4C~71.07 & Jun 24, 2013 - Nov 23, 2015 & 10 & 40.5 $\pm$ 11.8 & 2.7 $\pm$ 4.2 & -6.15 $\pm$ 15.5 \\
    PKS~1510-089 & May 30, 2013 - Jun 30, 2017 & 85 & -26.6 $\pm$ 20.5 & 4.9 $\pm$ 4.0 & 5.9 $\pm$ 51.7 \\
    4C~38.41 & May 31, 2013 - Sep 5, 2016 & 93 & -57.0 $\pm$ 16.1 & 6.6 $\pm$ 4.3 & 11.7 $\pm$ 43.0 \\
    3C~345 & Jun 8, 2013 - Sep 18, 2017 & 36 & 81.8 $\pm$ 14.0 & 4.9 $\pm$ 2.7 & 52.5 $\pm$ 30.0 \\
    4C~09.57 & May 31, 2013 - Sep 5, 2016 & 94 & -2.3 $\pm$ 17.0 & 10.1 $\pm$ 5.6 & -20.4 $\pm$ 45.6 \\
    BL~Lac & Jun 19, 2913 - Dec 1, 2023 & 1603 & 9.2 $\pm$ 7.2 & 10.7 $\pm$ 9.1 & 32.2 $\pm$ 37.6 \\
    3C~446 & Jun 23, 2013 - Aug 16, 2015 & 10 & -54.1 $\pm$ 19.0 & 15.3 $\pm$ 6.2 & -25.4 $\pm$ 26.4 \\
    4C~11.69 & Jun 23, 2013 - Oct 20, 2017 & 97 & -50.3 $\pm$ 13.3 & 4.4 $\pm$ 3.6 & 6.4 $\pm$ 39.8 \\
    3C~454.3 & Jun 21, 2013 - Jul 2, 2023 & 209 & -81.0 $\pm$ 19.5 & 4.3 $\pm$ 3.4 & -26.3 $\pm$ 45.5 \\
\hline
\end{tabular}
\tablefoot{\centering The column "Time span of observations" indicates the dates of the first and the last observation. Note that the observations do not cover the entire period indicated in the column, but rather intervals within that time span.
The column "Number of observations" indicates the number of measurements taken into account for each source.
The columns "PD" and "PA" refer to the median values over all the observations taken into account in this work, with their errors.
The jet directions defined as "Variable" indicate that different values have been used for different observations (see Table \ref{table:hsp_obs} for details).
Both jet direction and angle are defined between [-90°, 90°].}
\label{table:sample}
\end{table*}

\begin{table*}[h]
\caption{\centering IXPE observations of the HSP sample taken into account in this work, together with the previous publications used for this analysis.}
\centering
\begin{tabular}{c c c c c}
\hline\hline
\noalign{\vskip 0.3mm}
Source & \texttt{obsID} & Dates & Jet direction & Polarisation properties \\
\noalign{\vskip 0.2mm}
\hline
\noalign{\vskip 0.4mm}
\multirow{4}{*}{Mrk~421} 
   & 01003701 & May 4-6, 2022 & \multirow{3}{*}{\citet{weaver2022kinematics}} & \citet{di2022x} \\
   & 01003801 & Jun 4-6, 2022 & & \citet{di2023discovery} \\
   & 02004401 & Dec 6-8, 2022 & & \citet{kim2024magnetic} \\
   & 02008199 & Dec 6-22, 2023 & \citet{maksym2024two} & \citet{maksym2024two} \\
\noalign{\vskip 0.2mm}
\hline
\noalign{\vskip 0.4mm}
\multirow{10}{*}{Mrk~501} 
   & 01004501 & Mar 8-10, 2022 & \multirow{10}{*}{\citet{weaver2022kinematics}} & \multirow{6}{*}{\citet{chen2024x}} \\
   & 01004601 & Mar 27-29, 2022 & & \\
   & 01004701 & Jul 9-12, 2022 & &\\
   & 02004601 & Feb 12-14, 2023 & & \\
   & 02004501* & Mar 19-21, 2023 & & \\
   & 02004701 & Apr 16-18, 2023 & & \\
   & 03008801 & Apr 8-10, 2024 & & This work \\
   & 03008901 & Jul 7-9, 2024 & & This work \\
   & 03009001 & Aug 10-12, 2024 & & This work \\
   & 03009101* & Jan 3-5, 2025 & & This work \\
\noalign{\vskip 0.2mm}
\hline
\noalign{\vskip 0.4mm}
\multirow{4}{*}{1ES~1959+650} 
   & 01006201 & May 3-4, 2022 & \multirow{2}{*}{\citet{errando2024detection}} & \multirow{2}{*}{\citet{errando2024detection}} \\
   & 01006001$\star$ & Jun 9-12, 2022 & & \\
   & 02004801 & Oct 28-31, 2022 & \multirow{2}{*}{\citet{pacciani2025}} & \multirow{2}{*}{\citet{pacciani2025}} \\
   & 02250801 & Aug 14-19, 2023 & & \\
\noalign{\vskip 0.2mm}
\hline
\noalign{\vskip 0.4mm}
   1ES~0229+200 & 01006499 & Jan 15-Feb 1, 2023 & \citet{piner2018multi} & \citet{ehlert2023x} \\
\noalign{\vskip 0.2mm}
\hline
\noalign{\vskip 0.4mm}
   PG~1553+113 & 02004999 & Feb 1-9, 2023 & 90° $\dag$ & \citet{middei2023ixpe} \\
\noalign{\vskip 0.2mm}
\hline
\noalign{\vskip 0.4mm}
   PKS~2155$-$304 & 02005601 & Oct 27-Nov 7, 2023 & \citet{kouch2024ixpe} & \citet{kouch2024ixpe} \\
\noalign{\vskip 0.2mm}
\hline
\end{tabular}
\tablefoot{\centering The columns "Jet direction" and "Polarisation properties" refer to the works from which the values of the jet direction and the average PD, PA and errors were taken.
The observations reported with * were not taken into account for this analysis, as they were $3\sigma$ upper limits on PD and not detections.
The observation reported with $\star$ was originally divided into four equal time bins, of which we only used the third, as the other three were upper limits.
The value reported with $\dag$ comes from private communications. No error was indicated, therefore the value used was the average over the errors of the other jet measurements.
For the new IXPE observations of Mrk~501 (indicated with "This work"), see Table \ref{table:new_obs}.}
\label{table:hsp_obs}
\end{table*}

\section{New multiwavelength observations of Mrk~501}

\begin{table*}[h]
\caption{\centering Polarisation properties of Mrk~501 measured by IXPE during the new observations (2024, 2025) carried out during Cycle 1.}
\centering
\begin{tabular}{c c c c c c c}
\hline\hline
\noalign{\vskip 0.3mm}
\texttt{obsID} & Start time [MJD] & End time [MJD] & PD [\%] & PD error [\%] & PA [deg] & PA error [deg] \\
\noalign{\vskip 0.2mm}
\hline
\noalign{\vskip 0.4mm}
   03008801 & 60408.108 & 60410.176 & 17.4 & 2.8 & 124.7 & 4.6\\
   03008901 & 60498.836 & 60500.912 & 11.7 & 2.4 & 135.5 & 6.0\\
   03009001 & 60530.194 & 60534.295 & 11.2 & 2.9 & 137.2 & 7.5\\
   03009101 & 60678.163 & 60680.194 & $\leq$14.9 & - & - & -\\
\noalign{\vskip 0.2mm}
\hline
\end{tabular}
\tablefoot{\centering Start and end time are reported, together with the median polarisation degree and its error, as well as the median polarisation angle and its error.
The last observation was not a detection, but a $3\sigma$ upper limit, and therefore has no measurement on the PA and it was not included in the analysis.}
\label{table:new_obs}
\end{table*}

\begin{table*}[h]
\centering
\caption{\centering Median values of the polarisation properties measured during multiwavelength observations of Mrk~501, simultaneous to the new IXPE observations.
}
\begin{tabular}{c c c c c c}
\hline\hline
\noalign{\vskip 0.3mm}
\texttt{obsID} & Band & PD [\%] & PD error [\%] & PA [deg] & PA error [deg] \\
\noalign{\vskip 0.2mm}
\hline
\noalign{\vskip 0.4mm}
   \multirow{4}{*}{03008801} & 4 GHz & 1.8 & 0.2 & 175.3 & 2.6 \\
    & 10 GHz & 3.2 & 1.6 & 166.9 & 3.8 \\
    & 15 GHz & 1.4 & 0.2 & 161.6 & 2.5 \\
    & R & 1.8 & 0.2 & 175.3 & 2.6 \\
\noalign{\vskip 0.2mm}
\hline
\noalign{\vskip 0.2mm}
   \multirow{3}{*}{03008901} & R & 4.8 & 1.2 & 126.9 & 5.5 \\
    & V & 2.8 & 0.3 & 121 & 3.9 \\
    & B & 2.5 & 0.5 & 127.4 & 4.8 \\
\noalign{\vskip 0.2mm}
\hline
\noalign{\vskip 0.2mm}
   \multirow{4}{*}{03009001} & 225 GHz & 2.6 & 0.7 & 159.1 & 5.7 \\
    & R & 6.7 & 1.6 & 125.5 & 5 \\
    & V & 3.4 & 0.1 & 124.2 & 1.0 \\
    & B & 4.4 & 0.2 & 124.9 & 0.5 \\
\noalign{\vskip 0.2mm}
\hline 
\noalign{\vskip 0.2mm}
03009101 & R & 3.6 & 0.2 & 125.5 & 1.3 \\
\noalign{\vskip 0.2mm}
\hline 
\end{tabular}
\tablefoot{\centering Only the bands that had available measurements are reported.}
\label{table:new_obs_mwl}
\end{table*}

\end{appendix}

\end{document}